\documentclass[letterpaper]{article}
\usepackage{aaai}
\usepackage{times}
\usepackage{helvet}
\usepackage{courier}
\usepackage{graphicx}
\usepackage{subcaption}
\usepackage{float}
\usepackage{rotating}
\usepackage{amsmath}
\usepackage{placeins}
\usepackage{algorithm}
\usepackage[noend]{algpseudocode}
\usepackage{amsthm}
\usepackage{amsmath}
\usepackage{tikz}
\usepackage{url}

\newcommand{\madmax}{$\mu_{max}$ \ }
\newcommand{\madmaxns}{$\mu_{max}$}
\newcommand*\circled[1]{\tikz[baseline=(char.base)]{%
            \node[shape=circle,draw,inner sep=1pt, minimum size=4mm] (char)
            {#1};}}
            
\theoremstyle{definition}
\newtheorem{defin}{Definition}

\title{The Early Bird Catches The Term: Combining Twitter and News Data For Event Detection and Situational Awareness}

\author{Nicholas Thapen\qquad Donal Simmie \qquad Chris Hankin\\ Institute for 
Security Science and Technology,
Imperial College London}

\date{}

\newcommand{\specialcell}[2][c]{%
     \begin{tabular}[#1]{@{}l@{}}#2\end{tabular}}

\begin{document}
\frenchspacing
\maketitle

\begin{abstract}
\begin{quote}
Twitter updates now represent an enormous stream of information 
originating from a wide variety of formal and informal sources, much of which 
is relevant to real-world events. In this paper we adapt existing 
bio-surveillance algorithms to detect localised spikes in Twitter activity 
corresponding to real events with a high level of confidence. We then develop a 
methodology to automatically summarise these events, both by providing the 
tweets which fully describe the event and by linking to highly relevant news 
articles. We apply our methods to outbreaks of illness and events strongly 
affecting sentiment. In both case studies we are able to detect events 
verifiable by third party sources and produce high quality summaries.
\end{quote}
\end{abstract}

\section{Introduction}

Updates posted on social media platforms such as Twitter contain a great deal of
information about events in the physical world, with the majority of topics
discussed on Twitter being news related \cite{kwak2010twitter}. Twitter can therefore be used
as an information source in order to detect real world events. The content and 
metadata contained in the tweets can then be leveraged to describe the events 
and provide context and situational awareness. Applications of event detection 
and summarisation on Twitter have included the
detection of disease outbreaks \cite{aramaki11}, natural disasters such as
earthquakes \cite{sakaki2010earthquake} and reaction to sporting events
\cite{zubiaga2012towards}.

Using the Twitter stream for event detection yields a variety of advantages.
Normally in order to automatically detect real-world events a variety of
official and media sources would have to be tracked. These are usually published
with some lag time, and any system monitoring them programmatically would
require customisation for each source since they are not formatted in any
standard way.
Twitter provides a real-time stream of information that can be accessed via a
single API. In addition a rich variety of sources publish information to
Twitter, since it is a forum both for the traditional media and for a newer
brand of citizen journalists \cite{hermida2010twittering}. Tweets also contain metadata that can be mined for information, including location data, user-supplied
hashtags and user profile information such as follower-friend relationships. The
primary drawback of using Twitter is that it is an unstructured source that
contains a great deal of noise along with its signal. Tweets can be inaccurate as a result of rumour, gossip or active manipulation via spamming.

In this paper we apply existing bio-surveillance algorithms to detect candidate
events from the Twitter stream, employing customised filtering techniques to
remove spurious events. We then extract the terms from the event tweets which
best characterise the event and are most efficacious in retrieving related news.
These terms are used to filter and rank the most informative tweets for
presentation to the user along with the most relevant news articles.

Our techniques are evaluated using two case studies, both using a dataset of
geo-located tweets from England and Wales collected in 2014. The primary case
study is the detection of illness outbreak events. We then generalise our
techniques to events strongly affecting Twitter sentiment, such as celebrity
deaths and big sports matches.

In Section \ref{sec:rel-work} we discuss related work in the area of event
detection and situational awareness using Twitter. Sections \ref{sec:method} and
\ref{sec:results} outline our methodology and results. We then discuss our
conclusions in Section \ref{sec:conclusion}.

\section{Related Work}
\label{sec:rel-work}

Much of the work on event detection using social media has focused on using
topic detection methods to identify breaking news stories. Streaming document 
similarity measures \cite{petrovic2010streaming}, \cite{osborne2014real} and 
online incremental 
clustering \cite{becker2011beyond} have been shown to be effective for this 
purpose.

Other approaches have aimed to pick up more localised events. These have 
included searching for spatial clusters in tweets \cite{walther2013geo}, 
leveraging the social network structure \cite{aggarwal2012event}, analysing the 
patterns of communication activity \cite{chierichetti2014event} and identifying 
significant keywords by their spatial signature \cite{abdelhaq2013eventweet}.

In the field of disease outbreak detection efforts have mostly focused on 
tracking levels of influenza by comparing them to the level of self-reported 
influenza on Twitter, in studies such as 
\cite{broniatowski2013national} and \cite{li2013early}. Existing disease 
outbreak detection algorithms have also been applied to Twitter data, 
for example in a case study \cite{diaz12} of a non-seasonal
disease outbreak of Enterohemorrhagic Escherichia coli (EHEC) in Germany. They
searched for tweets from Germany matching the keyword ``EHEC'', and used the daily
tweet counts as input to their epidemic detection algorithms.
Using this methodology an alert for the EHEC outbreak was triggered before
standard alerting procedures would have detected it. Our study uses a
modified and generalised version of this event detection approach.

Diaz-Aviles \textit{et al.} also attempted to summarize outbreak events by 
selecting the most
relevant tweets, using a customized ranking algorithm. Other studies which have summarised 
events on Twitter by selecting the most relevant tweets include \cite{zubiaga2012towards} and
\cite{long2011towards}.

There has been less related work on linking or substantiating events detected
from Twitter with traditional news media. One study \cite{abel2011semantic}
analysed various methods of contextualizing Twitter activities by linking them
to news articles. The methods they examined included finding tweets with
explicit URL links to news articles, using the content of tweets, hashtags and
entity recognition. The best non-URL based strategy that they found was the
comparison of named entities extracted from news articles using OpenCalais with
the content of the tweets.

\section{Methodology}
\label{sec:method}

\subsection{Problem Definition}

Our definition of a real-world event within the context of Twitter is taken from
\cite{becker2011beyond}, with the exception that we have added a concept of event location.  

\begin{defin}
(\textbf{Event}) An event is a real-world occurrence \textit{e} with (1) an associated
time period $T_e$ and (2) a time-ordered stream of Twitter messages $M_e$, of substantial volume,
discussing the occurrence and published during time $T_e$. The event has a location $L_e$ where it took place, which may be specific or cover a large area, and the messages have a set of locations 
$L_{M1}$,...,$L_{Mn}$ which they were sent from.
\end{defin}

When given a time-ordered stream of Twitter messages M, the event detection problem is therefore one of identifying the events $e_1$,...,$e_n$ that are present in this stream and their associated time periods $T_e$ and messages $M_e$. It is also valuable to identify the primary location or locations $L_{Mi}$
that messages have originated from, and if possible the event location $L_e$.
The situational awareness problem is one of taking the time period $T_e$ and 
messages $M_e$ and producing an understandable summary of the event and its 
context.

\subsection{Overview}

Our approach to the event detection problem incorporates location by detecting 
deviations from baseline levels of tweet activity in specific geographical 
areas. This allows us to track the 
location of messages relating to events, and in some cases determine the event 
location itself.
We break down the problem by defining classes of events which we are interested in and formulating a set of groups of keywords which describe each class. In this paper we have examined two distinct classes:
\begin{itemize}
\item Outbreaks of symptoms of illness, such as coughing or itching
\item Events triggering emotional states, such as happiness or sadness
\end{itemize}

We track the number of tweets mentioning each keyword in each of our areas and use modified bio-surveillance algorithms to detect spikes in activity which we can classify as events. 

Initially we designed the system with health symptom event detection as the
primary use case. This led to a system design focused around keywords and
aliases for their keywords, since a limited range of illness symptoms
characterises most common diseases and the vocabulary used to describe these
symptoms is also relatively limited. After several iterations of this approach
we noted that it could be viable as a general event detection and situational
awareness method, so we added another event class, emotion-based events, to
test out the feasibility of the general approach.

Our situational awareness approach is based on identifying terms from the event tweets which
characterise the events and using them to retrieve relevant news articles and
identify the most informative tweets.
The news search uses metrics based on cosine similarity to ensure that searches
return related groups of articles.

\subsection{Architecture}

\begin{figure}
	\centering
	\includegraphics{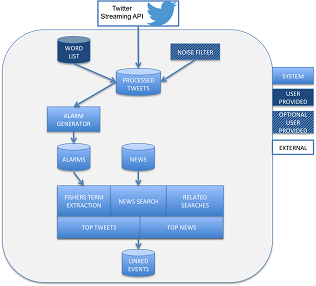}
	\caption{\textbf{Event Detection and Situational Awareness architecture:}
		To apply to a new example a
		user needs to provide a keyword group list and optionally a noise filter to remove
		tweets that do not strictly match the criteria of interest.}
	\label{fig:arch-diag}
\end{figure}

The general approach can be described by the architecture in Figure
\ref{fig:arch-diag}. Every new event class requires a list of
keyword groups. Optionally a domain specific data pre-processing step
can also be included. For example in the health symptom case we employ a machine learning classifier to remove noise (those tweets not actually concerning health). These are the only two aspects of the design
that need to be altered to provide event detection and situational awareness to a
new problem domain.

\subsection{Event Classes}

We now go into a more detailed explanation of our event classes and how we formulated the keyword groups. Each keyword group consists of a primary keyword which is used to identify the group, e.g. vomit, and a number of aliases that expand the group, e.g. throwing up, being sick, etc.

\label{sec:keywords}
\subsubsection{Illness Symptoms}

To build up a list of symptoms and related keywords we searched
Freebase for \texttt{/medicine/symptom}. Each of these symptoms is defined as a 
primary keyword. They are returned with a list of aliases that are
used as related keywords.

The next step in creating a symptom list was to filter these symptoms by their
frequency in the Twitter data, since only those words actually used on Twitter are of interest. All symptoms with less than 10 mentions in the Twitter data were
removed from this candidate list. This excluded a large proportion of symptoms,
reducing the set from ~2000 to 200. 

We further limited the set by removing symptoms not related to infectious diseases.
We also added primary keywords and aliases for some common conditions such as hayfever
and flu. This step resulted in 46 symptom groups.

\subsubsection{Emotion States}

For a list of emotion states and associated keywords we used the work of
Shaver \textit{et al.} . They conducted research \cite{shaver1987emotion} to 
determine
which sets of words were linked to emotions and how these cluster together. We took the six
basic emotions identified in the work as primary keywords: \textit{love, joy, surprise, sadness, anger and
fear}. Shaver's work associated each of these with a list of terms to form a
tree. We took the terms from lower leaves on the tree for each emotion as our
alias sets (See Table \ref{tab:senti-keywords} for examples). The only alteration we made was that after some initial
analysis we discovered that the term ``happy'' from the ``joy" category was a very
strong signal of special events such as Valentine's Day, Mother's Day and
Easter. It was also very often used on a daily basis due to people offering
birthday greetings. We therefore separated ``happy" into its own category
separate from ``joy".

In addition we employed SentiStrength \cite{thelwall2010sentiment}, a sentiment analysis 
tool, to classify our tweets into positive and negative emotional sentiment. We took those
classified as being very positive and very negative as additional emotion states.

\begin{table}
	\centering
	\begin{tabular}{ll}
		\hline
		Keyword & Aliases \\
		\hline
		surprise & amazed, astonished, surprised... \\
		sadness & depressed, unhappy, crying... \\
		joy & \specialcell{glad, delighted, pleased...\\} \\
		\hline
	\end{tabular}
	\caption{\textbf{Selected emotion keyword groups and some of their aliases:} keyword groups contain a
		primary keyword and aliases (taken from Shaver \textit{et al.} ).}
	\label{tab:senti-keywords}
\end{table}

\subsection{Data Collection}
\label{sec:data}
Using Twitter's live streaming API we collected geo-tagged tweets between 11th
February 2014 and 11th October 2014. Tweets were collected from within a
geographical bounding box containing England and Wales. Retweets were excluded
due to our focus on tweets as primary reports or reactions to events. This
resulted in a data-set of 95,852,214 tweets from 1,230,015 users.  1.6\% of 
users geo-tag their tweets \cite{leetaru13}, so our data is a limited sample of 
the total tweet volume from England and Wales during this period. We chose
to use only geo-tagged tweets since they contain metadata giving an accurate location for the user. This allows us to locate each tweet within our geographical model.

\subsection{Location Assignment}
\label{sec:location}

\begin{figure*}
\centering
\includegraphics{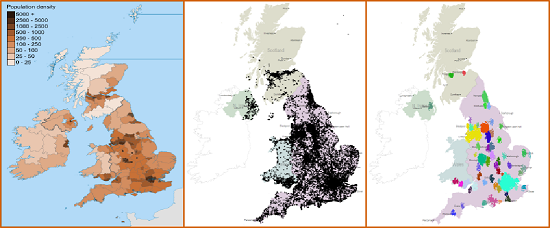}
\caption{UK population density (left) compared to a sample of geo-located 
tweets (centre) and the clusters found (right). Note that only clusters located 
in England and Wales were used in this study.}
\label{fig:cluster-pop-comp}
\end{figure*}

Our methodology relies on the collection of baseline levels of tweet activity in
an area, so that alarms can be triggered when this activity increases. We
therefore amalgamated the fine-grained location information from the geo-coded
tweets by assigning them to broader geographical areas.
We used a data driven approach to generate the geographical areas rather than using administrative areas such as towns or counties. This
technique allowed us to select only those areas with a minimum level of tweet activity, and also did not require any additional map data. It would therefore be be reusable for any region or country with a sufficient level of Twitter usage.

We began by viewing a sample of the collected tweets as geo-spatial points. Viewed on a map
these clearly clustered in the densely populated areas of England and Wales. We therefore decided to use a clustering algorithm on these points in order to separate out areas for study. We employed the
Density-Based Spatial Clustering of Applications with Noise
(DBSCAN) algorithm \cite{ester1996density} for clustering, as this does not 
require 
\textit{a priori} knowledge of the number of clusters in the data. The features 
provided to DBSCAN were the latitudes and longitudes of the tweets.

The clusters produced by the algorithm matched the most populated areas, corresponding to the
largest cities or towns in the UK as shown in Figure \ref{fig:cluster-pop-comp} . They also separated most cities into distinct
clusters (a notable exception being the conglomeration of Liverpool and
Manchester). In total 39 clusters were created for England and Wales and each was given an ID and a
label. We then created a convex hull around each cluster, providing a polygon
that can be used to check whether a point is in the cluster or outside it.
Points outside all of the clusters were assigned to a special 'noise' cluster,
and not included in the analysis. Overall 80\% of tweets were assigned to specific clusters and the remainder to noise, giving us good coverage of geo-tagged tweets using our cluster areas.

\subsection{Tweet Processing}
\label{sec:processing}

As tweets are received by our system they are processed and assigned to the
symptom and emotion state classes via keyword matching. They are assigned a location by checking whether they fall into one of our cluster areas.

For the illness symptoms we introduce a noise removal stage at this point. It is particularly relevant for this class of events because there are many fewer tweets relating to illness than showing emotion states. This means that the signal is more easily blocked out by random noise. To remove noise we construct a machine learning classifier with the aim of removing tweets containing alternative word usages or general illness discussion rather than reporting of illness events. The classifier we use is a linear SVM trained on a semi-supervised cascading
training set \cite{sadilek2012predicting}. This
classifier uses the LibSVM \cite{CC01a} library, and achieves a classification
accuracy of 96.1\% on a test set of manually classified tweets.

The number of tweets assigned to each class in each area are then saved on a daily basis. These counts are
first normalised to take account of Twitter's daily effect pattern, which shows
more tweeting on weekends than weekdays.
Event detection is run daily since we are attempting to pick up temporally coarse-grained events. Disease outbreaks take weeks to develop, and events that shift
public sentiment or emotion will generally take hours or days to unfold.

\subsection{Detecting Events}
\label{sec:ewd-method}

Our event detection methodology leverages considerable existing syndromic surveillance research by
using an algorithm designed and developed by the Centers for Disease Control and Prevention (CDC), the Early Aberration Reporting System
(EARS) \cite{hutwagner03}.

\begin{defin}
(\textbf{Alarm}) An \textit{alarm} is an alert produced by the first stage of our event detection system. The alarm has an associated symptom and location. It also has a start and end date, and associated tweet counts for each date within this period. When certain criteria are met an alarm is deemed to be an event.
\end{defin}

We employ the C2 and C3 variants of EARS. These algorithms operate on a time series of count data, which in our case is a count of daily symptomatic tweet activity. The C2 algorithm uses a sliding seven day baseline, and signals an
alarm for a time $t$ when the difference between the actual count at $t$ and the
moving average at $t$ exceeds 3 standard deviations. The C3 algorithm is based on
C2, and in effect triggers when there have been multiple C2 alarms over the
previous 3 days.

These C2 and C3 candidate alarms are then grouped together so that alarms for
the same keyword set and area on consecutive days are treated as a single alarm.
An alarm is therefore made up of one or more days, each with an observed count
of tweets. 

Some of our Twitter count time series data is zero-skewed
and non-normal, since the number of geo-tagged users reporting illness can be
low. The number of standard deviations from the mean used in the C2 and C3
algorithms can be an unreliable measure of central tendency in those circumstances.
Hence to determine how far above general baseline activity an observed count is
we employ the median of the series to date and the Median Absolute Deviation (MAD) to
produce a new metric of alarm severity. The number of Median Absolute Deviations
from the median, $\mu$, gives a comparable figure across alarms as to how sharp
a rise has been over expected levels. This figure is produced from the following
equation: %%%%%%%%% No gap between text and eqn environment
\begin{equation}
\mu = (observation-median)/MAD
\end{equation}

We then find the highest metric for an alarm, \madmax, by finding the highest value of $\mu$
within the observations making up the alarm.%%%%%%%%% No gap between text and eqn environment
\begin{equation}
\mu_{max} = \operatorname*{arg\,max}_\mu (\mbox{\textit{observations in alarm}})
\end{equation}

The \madmax is the primary statistic which we use to determine which events are
real and which have just been generated by random noise. Details of the threshold value
which we use for this and how we selected it are contained in Section 
\ref{sec:results}.

Another statistic which we employ in order to filter out noise is the tweet-user
ratio. This is the ratio of tweets in an event to that of distinct users
involved in an event. A high value of this statistic would imply that some users
have tweeted a large number of times across a short time period, which is an
indication that they may be spammers and that the alarm is spurious.

In summary, we use the output from EARS to produce alarms. We filter the alarms to a set of high likelihood events by using the \madmax and tweet-user ratio parameters.
		
\subsection{Situational Awareness}	
\label{sec:sa-method}
Once an event has been identified our next objective is to automatically provide
additional context for it, which may provide an explanation of the
underlying cause. A human interpreter could achieve this by reading all of the
tweets and synthesizing them into a textual explanation, which might be some
text such as ``People reacting to the death of Robin Williams''. We do this in
two main ways: by providing the most representative tweets from those that
triggered the alarm, and by linking to relevant news articles. The steps
involved in the Terms, News and Tweets (TNT) Event Summarisation process are detailed in
Algorithm \ref{alg:ate}. The steps and terminology are then explained in more detail.

\begin{centering}
\begin{algorithm}[htb]

\caption{Terms, News and Tweets (TNT) Event Summarisation}
\label{alg:ate}
\algrenewcommand{\alglinenumber}[1]{\scriptsize\circled{#1}}
\begin{algorithmic}[1]
\State Fetch gist tweets and baseline tweets
\If{$\left\vert{\text{gist tweets}}\right\vert < 30$} \State
Do not attempt to summarise event
\Else
	\State Extract unigrams and bigrams appearing in at
	least 5\% of the gist tweets
	\ForAll{ngrams extracted}
		\State Perform Fisher's Exact Test to determine whether ngram is significantly more likely to appear in gist than baseline
	\EndFor
	\For{Top 2 most significant unigrams and bigrams and the primary keyword}
		\State Search news database using ngram for the alarm's date range and return
		the top 10 documents \State Compute PCSS for documents returned
	\EndFor
	\For{ngrams with PCSS values above threshold}
		\State Compute title similarity PCSS between ngram documents and those for each other ngram
		\State Good search terms $\gets$ term with title similarity PCSS above
		threshold
	\EndFor
	\State Good articles $\gets$ documents returned from good search terms
	\State Filtered tweets $\gets$ tweets containing a good search term
	\State Rank good articles by cosine similarity to average vector of good
	news articles 
	\State Rank filtered tweets by cosine similarity to average vector of
	filtered tweets
\EndIf
\end{algorithmic}
\end{algorithm}
\end{centering}

\circled{1} The first step is to retrieve the relevant tweets from
the processed tweet and alarm databases. Tweets are fetched for both the alarm gist and from a
historical baseline. 
\circled{3} We discard those events with fewer than 30 tweets as we found that they did not contain sufficient data to produce good summarisation results. 

\begin{defin}
(\textbf{Gist}) The gist consists of the tweets for the time period of the event which match the event's
keyword group and area.
\end{defin}

\begin{defin}
(\textbf{Baseline}) The baseline consists of the tweets for the same keyword group and area as an event from the 28 days prior to that event.
\end{defin}

\circled{5} The next task is to find unigrams and bigrams that are more
prevalent in the gist than in the baseline. These are likely to come
from tweets discussing the event and will thus be characteristic of the event.
We first extract the most common unigrams and bigrams from both sets of tweets,
after removal of stopwords. Our list of stopwords includes a standard list, plus
the 200 most frequent words from our tweet database. We select all non-stopwords
that appear in at least 5\% of the tweets.

\circled{7} We then do a Fisher's Exact Test to determine which of the common
unigrams and bigrams in the gist appear significantly more frequently ($\alpha < 0.05$)
here than in the baseline set. Our candidate terms are the top two most
significant unigrams and bigrams. We select the top two as this was found to give the best results on our test examples. To this set we append the primary keyword that
triggered the alarm.

\circled{9} Using these candidate terms we then perform a search on Google for
documents published in the United Kingdom during the time period of the alarm. Due to
Google's Terms of Service this step was performed manually. A fully automated
system would replace this step with a search of a news database, which could be
created by pulling down news articles from RSS feeds of major content providers.

\circled{10} We take the first 10 documents retrieved for each search term,
remove stopwords and apply stemming using a Lancaster stemmer. We then convert
each document into a Term Frequency/Inverse Document Frequency (TF/IDF) vector. In order to determine whether the search
term has retrieved a coherent set of related documents we define a metric based
on cosine similarity, the Pairwise Cosine Similarity Score (PCSS).

\begin{itemize}
\item The \textbf{Pairwise Cosine Similarity Score} of a group of TF/IDF vectors is calculated by taking the cosine similarity between each pair of vectors and adding them to a set. The standard deviation of this set is subtracted from its mean to form a score. 
\end{itemize}

The PCSS rewards articles which are similar and penalises any variance across
those article similarities, this reduces the effect of some articles being
strongly related in the document set and others being highly unrelated. Any term
which retrieves a set of documents with a score below a threshold value
(determined by a parameter selection process detailed in section
\ref{sec:results}) is not considered further.

It is possible for a search term to hit on a coherent set of documents purely by
chance, perhaps by finding news articles related to another event in a different
part of the world. In order to guard against this we institute another check to
ensure that the set of documents returned from a search term is sufficiently
closely related to the set returned from at least one other search term.

\circled{12} In order to perform this check we compare the titles of the
articles returned from the two different searches using a similar process to our earlier document comparison. We found it more effective to compare titles than whole documents, since sets of documents with similar
topics can contain similar language even for fairly unrelated search terms. For
example the terms ``ebola'' and ``flu'' will both return health-related documents
containing similar language, but we would not wish to say that these search
terms are related. To convert the titles to TF/IDF vectors we remove stopwords
but do not apply stemming. Since the titles are so short we include all unigrams, bigrams and
trigrams in the vector representation. We then compute a PCSS between the two
document sets, pairing each document in the first set with each in the second
and vice versa. \circled{13} A search term must be related to at least one other
term for it to be used going forward. 

\circled{14} Once TNT has identified good search terms we then return the news
articles fetched using those terms. \circled{16} In order to rank the top news
articles for a search we take the average TF/IDF vector and then rank the
articles by cosine similarity to this average vector. We return the top ranked
articles from each search term.

\circled{17} In order to return the most explanatory tweets we find the
gist tweets that contain at least one of the good search terms. We then convert these into
TF/IDF vectors and compute the average vector. The tweets are then ranked in the
same way, by cosine similarity to the average vector, and we return the top 5
tweets.

\section{Results}
\label{sec:results}

There are three individual components to our event detection and situational
awareness platform that require evaluation: 

\begin{enumerate}
  \item Event detection
  \item Situational awareness
  \begin{enumerate}
  	\item Linkage of relevant news articles
  	\item Ranking most informative tweets
  \end{enumerate}
\end{enumerate}

\subsection{Example Cases}
\label{sec:eval-alarms}

To effectively evaluate all of these components required a varied set of example
events and alarms. These were used in order to choose values for our threshold 
parameters. We compiled an initial set of 13 
focus examples. These
were taken from events that the authors knew had happened in the evaluation time
period and from those alarms in our dataset with low and high values of \madmaxns. The event ID which
will be used to refer to these events is composed of the
first two letters of the event keyword followed by a 1-2 letter area code. The 
final part of the ID is the day and month of the event start date.

The focus examples were used to find sensible values that separated the 
high-confidence events from the low-confidence events. The most important 
threshold parameter in
the context of the event detection is the \madmax figure which measures the 
deviation of the alarm counts from the median level. 
Examining the distribution of the number of alarms for each value of \madmax 
revealed that it started to tail off sharply at $\mu_{max} \geq 5$. We 
therefore took this as a value to segment additional test examples, drawing ten 
more at random with a \madmax less than 5 and ten with a \madmax greater than or equal to 5.

\subsection{Event Detection Evaluation}
\subsubsection{Method}
\label{sec:eval-method}

It is difficult to provide a completely automated evaluation procedure for
detecting previously unknown events. Diaz \textit{et al.} used the time to
detection on a known outbreak as their evaluation criterion \cite{diaz12}. In our
case we do not know \textit{a priori} that these are genuine outbreaks or
events. Hence we need to make an assessment of the alarms produced to see what
they refer to and if there is a way of externally verifying that they are genuine events.
For all 33 of the selected alarms the authors read the tweets and determined whether they described a real world event. The
coders found 26 YES answers, 5 NO answers and 2 DISAGREED answers, producing a
94\% agreement. Where an event was present they wrote a short summary,

For external verification of events two different methods were used, depending
on whether the event was symptom-related or emotion-based. For symptom related
events the activity spike was checked against official sources for the same time
period. The General Practitioner (GP) in hours bulletin for England and Wales
\cite{gp-in-hours-2014} was used and an event was deemed verified if the symptom
exhibited an increasing trend for that period. This detail is noted in the
summary document produced by Public Health England for that reporting period.
Emotion-based events were verified by checking if there were any articles (via
Web search) that could corroborate the cause of the event (as given by the summary).

We manually investigated all examples from the initial focus set and found
initial parameters for the score functions in our algorithms that worked reasonably
well. These provided possible ranges of values which were evaluated
more systematically over the entire alarm set. For event detection we evaluated which alarms were flagged as events by the system for each parameter value against whether those events were externally verifiable. The final evaluation for
all algorithms contains all 33 of the alarms in both sets, not just the twenty
expanded 'test' examples. 

\subsubsection{Results}
\label{sec:event-eval}

To determine if an alarm is an event that we should be concerned about we consider
two properties of the alarm. The first is the tweet-user ratio. This provides
a naive spam filter, as when this is high an alarm is mostly caused by one
user tweeting multiple times. From exploratory testing we found a value of $1.5$
separated our spam and genuine alarms very well, leaving only a small number of
alarms with large tweet sets and some spam. The spam detection problem should be
straightforward and will be addressed more completely in future work.

The second figure which gives the strength of the activity above the usual
baseline is the \madmax figure. This is the essence of the modified EARS
algorithm and the value of this figure should generally separate events from
non-events.

\begin{table*}
\centering
\begin{tabular}{llllllllll}
  \hline
  	ID         & Event & \madmaxns & Keyword & Node & ID         & Event & 
  	\madmaxns & Keyword & Node \\
	\hline
	SAL-11-08  & YES   & 20  & Sadness & London & HFB-10-04  & YES   & 5 & 
	Hayfever    &	Birmingham	 \\ 
	HFM-01-06  & YES   & 19   & Hayfever &	Manchester   & 
	VOL-20-04  & YES   & 5 & Vomit	 &	London	 \\ 
	SAL-07-04  & YES   & 14 	 & Sadness	 &	London	 & SAC-05-05  & YES   & 
	5 & Sadness     &	Cardiff	\\ 
	FEL-18-07  & YES   & 13 & Fear	 &	London		 & HFL-04-07  & NO    & 5 & 
	Hayfever    &	London	 \\ 
	ASL-02-04  & YES   & 12 & Asthma	 &	London		 & FLB-23-09  & NO*    
	& 5 & Flu	     &	Birmingham	\\ 
	FLP-06-10  & YES   & 11 & Flu	     &	Portsmouth  	 & 
	VPBR-10-05 & YES   & 4 & VeryPos &	Bristol	 \\ 
	HAM-02-04  & YES   & 9  & Happy	      &	Manchester 	 & 
	FRL-30-05  & YES   & 4 & Fever	      &	London	\\ 
	HAM-18-04  & YES   & 9  & Happy	      &	Manchester	 & 
	FLM-19-09  & YES   & 4 & Flu	      &  Manchester \\ 
	SAL-08-07  & YES   & 8   & Sadness	 &	London	 	 & VOL-22-02  & NO    & 
	3 & Vomit	 & 	London	\\ 
	HALE-01-08  & YES   & 8  & Happy	      &	Leeds 	 & 
	HFB-29-04  & NO    & 3  & Hayfever    &	Birmingham	  \\ 
	HFL-14-05  & YES   & 7  & Hayfever    &	Leeds	 	 & JONO-23-02  
	& YES   & 2 & Joy	      & Norwich	  \\ 
	SUN-29-08  & YES   & 7  & Surprise &  Newcastle		 & HEM-06-03  & NO    & 
	2 & Headache    &	Manchester \\ 
	ITL-08-06  & YES   & 6   & Itch	      &	London		 & SUC-23-05  & NO    & 
	2  & Surprise    &	Cardiff	  \\ 
	SAB-09-06  & YES   & 6  & Sadness     &	Birmingham		 & SUL-16-08  & 
	NO    & 1 & Surprise    & London	 \\ 
	HABE-01-03 & YES   & 5   & Happy	      &	Bridgend	 & FEBR-17-04 & 
	NO    & 0 & Fear	      & Bristol	 \\ 
	SAL-21-03  & YES   & 5  & Sadness	 &	London		 & STL-26-08  & NO    & 
	0  & Sore Throat & London	\\ 
	HFC-09-04  & YES   & 5  & Hayfever    &	Cardiff		 &            &       
	&   \\ 
  \hline
   
\end{tabular}
\caption{\textbf{Evaluation set of events:} showing whether they were 
externally verifiable and their \madmax value. *Note: this event not confirmed 
by the GP in hours report of that
week. However, the following week showed an increase and it is possible that
social media detected increased Influenza activity before this was
confirmed by GP visits.}
\label{tab:mad-max}
\end{table*}

The criterion for selecting the best threshold for \madmax is context dependent. We have used
the balanced F1 measure for this scenario as that is a fair representation of
both precision and recall. The classification success and error types are:

\begin{itemize}
\item \textbf{True positive}: instances at or above the threshold that are
verified events
\item \textbf{False positive}: instances at or above the threshold that are
not verified events
\item \textbf{True negative}: instances below the threshold that are not verified events
\item \textbf{False negative}: instances below the threshold that are verified events
\end{itemize}

\begin{figure}
\centering
\includegraphics[scale=0.5]{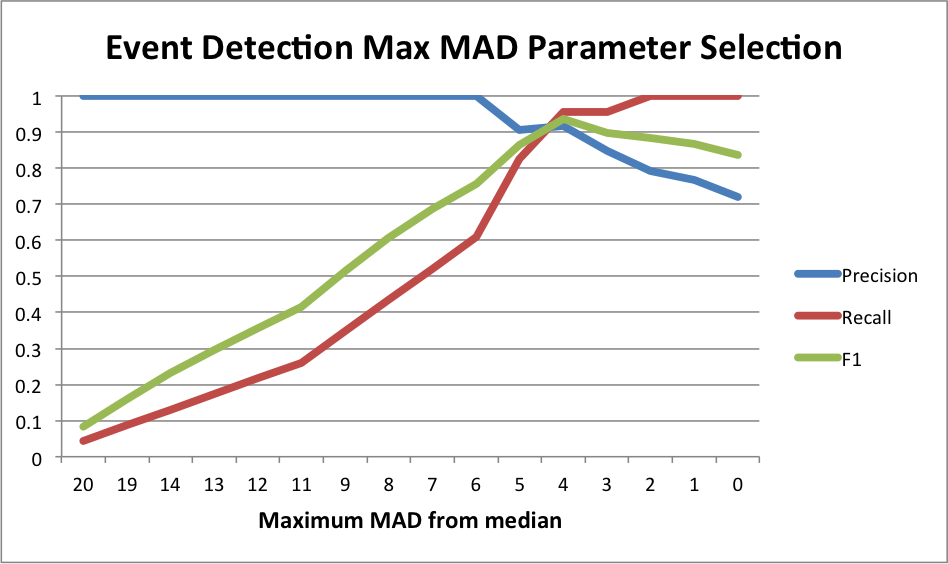}
\caption{\madmax event detection parameter selection}
\label{fig:madmax-event-detection}
\end{figure}

The precision, recall and F1 values for all the tested values of \madmax are
displayed in Figure \ref{fig:madmax-event-detection}. The maximum F1 value,
$0.9362$, is observed at $\mu_{max} \geq 4$, so this is a well balanced threshold and the
recommended parameter. Those seeking higher confidence events (willing to accept
that some events may be missed) could use a value of 6 for this parameter which yields a 
precision of 1. The maximum observed recall value is at the minimum parameter
value and is not very informative. Essentially it says that everything is an
event and hence does not produce any false negatives. 

In summary the event
detection mechanism based on the EARS C2 and C3 algorithms with the addition of
the \madmax and tweet-user ratio was found to perform well at detecting events that could be
externally verified as genuine. The recommended \madmax parameter (4) produced a good balance of precision and recall in our sample set. It must be noted however that we cannot gain a true picture of the overall recall of the system, since we have no way of analysing the number of genuine events that were not picked up. 

\subsection{Situational Awareness Evaluation}
\label{sec:sa-eval}

Both situational awareness components were evaluated. Firstly the news linkage 
was tested to see whether relevant news was retrieved for the sample events. As 
part of this analysis we compared our method of extracting informative search 
terms (the TNT algorithm) with a comparable automated technique. Secondly the 
tweet ranking was validated to determine whether highly ranked tweets 
effectively summarised the events.

\subsubsection{Comparative News Linkage Evaluation}
\label{sec:news-link-eval}

The news linkage component works by selecting good search terms for articles 
based on the TNT algorithm. Within this there is a term extraction step to 
generate search terms, and then a filtering step using PCSS to remove terms 
which retrieve unrelated sets of articles. We iterate over different threshold 
values for the PCSS score to find the optimum, using an F0.5 measure as the 
evaluation criterion. F0.5 was selected because precision was judged to be more
important than recall in this setting. As a further evaluation we compare the 
results of replacing our term extraction algorithm with Latent Dirichlet 
Allocation (LDA). LDA is a popular topic modelling technique
that extracts sets of terms characterising each topic in a group of documents.
The success and error types used to compute the F0.5 measure are:

\begin{itemize}
\item \textbf{True positive}: relevant news returned for newsworthy event
\item \textbf{False positive}: news returned for an event with no genuine news
\item \textbf{True negative}: no news returned for an event with no genuine news
\item \textbf{False negative}: no news returned for newsworthy event
\end{itemize}

\begin{figure}[htb]
\centering
\begin{subfigure}{.45\textwidth}
  \centering
  \includegraphics{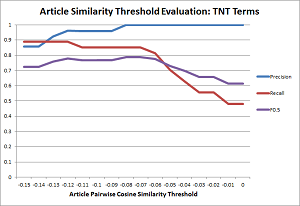}
  \caption{News linkage accuracy from Terms, News, Tweets terms}
  \label{fig:ate-results}
\end{subfigure}\\%
\begin{subfigure}{.45\textwidth}
  \centering
  \includegraphics{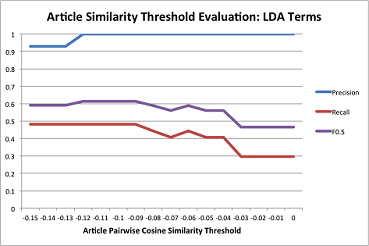}
  \caption{News linkage accuracy from Latent Dirichlet Allocation terms}
  \label{fig:lda-results}
\end{subfigure}
\caption{\textbf{Comparison of TNT and LDA event term extraction methods for
linking social and news media}}
\label{fig:news-linkage-results}
\end{figure}

The evaluation is presented in Figure \ref{fig:news-linkage-results} as well as
the different levels of article PCSS that were iterated over to find the maximum
F0.5 value in a step-wise procedure. It is clear from these images that the TNT
algorithm has a higher F0.5 at all tested values of the article PCSS, due to 
its higher recall. The outcome of the parameter selection process was that a 
PCSS threshold of $-0.08$ produced the best results. Using this value the F0.5 
was 0.79, showing that our system was successful in retrieving relevant news 
for the sample events.

Selecting top ranked relevant news articles is one part of our situational
awareness contribution. The second is the selection of tweets that provide a
representative summary of an event.
 
\subsubsection{Top Ranked Tweets Evaluation}
\label{sec:ranked-tweets-eval}

We select the summary tweets by choosing the top 5 tweets ranked by
calculating the maximum cosine similarity between an average tweet TF/IDF vector
and all tweets in the candidate set. This tweet set can be: 1) all tweets in the
gist, 2) those filtered by selecting the extracted terms or 3) those
from the filtered term set, that is, the extracted term set less any that don't
have a good news match. 1) is always available and is labelled the \textit{Gist
Top Tweets} (GTT).
If terms have been found to be significantly different in frequency from the
baseline then set 2) is available for use and if terms from that set have good
news matches then set 3) can be used. The \textit{Summary Top Tweets} (STT) are
from set 3) if it exists and fallback to set 2) if the good news match terms are not
available. If no terms were found to be significantly different from the
baseline then only the GTT is available.

We have employed two evaluations for the tweet ranking exercise: comparison to
human-coded event explanation and comparison between GTT and STT. The 
human-coded event explanations were created by both authors after reading 
through all of the tweets linked to each event. There were 26
alarms that had an identifiable cause. The tweet ranking match (to
human-coded event assessment) performance is presented in Table 
\ref{tab:stt-matching}. The tweets were considered a full match if a human 
summary of the 5 top ranked tweets would match the human-coded event 
explanation for the 
whole set of tweets.

\begin{table}
\centering
\begin{tabular}{lr}
  \hline
  Match   & Count \\ \hline
  Full    & 21 \\
  Partial & 2 \\
  No      & 3 \\
  \hline
  \end{tabular} 
\caption{\textbf{STT tweet ranking evaluation:}
The STT tweet summary fully matched the human-coded event summarisation in 21
cases. This yields a full match fraction of $0.81$.}
\label{tab:stt-matching}
\end{table}

The partial matches were: \texttt{FRL-30-05} (Fever: London, May) and
\texttt{FLP-06-10} (Flu: Birmingham, October). These events had more than one
explanatory cause. Currently our algorithms work best in the single event case. 
The three cases that did not match were: \texttt{JONO-23-02} (Joy:
Norwich, February), \texttt{STL-26-08} (Sore throat: London, August) and
\texttt{SUN-29-08} (Surprise, Newcastle, August). The coders disagreed as to
whether \texttt{STL-26-08} was actually an event. The remaining two examples were not
summarised well by the significant tweets as they both exhibited high disparity
in terms used to describe a contextually related event and \texttt{SUN-29-08}
also included a number of spam tweets that distorted the results of TNT.

The second evaluation for the tweet ranking exercise was a comparison between
the GTT and the STT. A qualitative assessment of the tweets led to the
conclusion that STT tweets were better in 11 out of 33 cases and there was no
significant difference between the two for 21 cases out of 33. In one case,
\texttt{FLP-06-10}, the GTT included a mention of ``flu jab'' (one of the
manually selected terms) which the STT did not include. Hence the STT provides
an improvement over ranking based off the alarm tweets in 1/3 instances.

\subsection{Notable Examples Discussion}
\label{sec:examples}

\begin{table*} \centering
\begin{tabular}{lll}
  	\hline
	ID         & TNT Terms   & LDA Terms \\
	\hline
	\texttt{JONO-23-02}  & joy, enjoy & enjoy, glad, loss \\
	\texttt{ASL-02-04}  & asthma, air pollution, smog, pollution & asthma, smog, pollution, attack air \\
	\texttt{VOL-20-04}  & vomit, chocolate, easter               & chocolate, 
	eaten, easter, vomit, headache \\
	\texttt{SAL-11-08}  & sadness, robin williams, sad news, robin, williams      
	& sad, robin, williams, rip, riprobinwilliams \\
  	\hline 
\end{tabular}
\caption{\textbf{Example cases and the terms extracted for them:} top terms
selected either by TNT or LDA.}
\label{tab:example-terms}
\end{table*}

We now discuss four example events that highlight the strengths and limitations 
of our approach. These examples are listed in Table \ref{tab:example-terms}.

The first example case is \texttt{JONO-23-02}. From a reading of the
tweets there were definitely some relating to a single event: Norwich
City Football Club beating Tottenham Hotspur Football Club $1-0$ in a football
match.
Both TNT and LDA term extraction failed to find terms representative of this event. This was due to
the disparity of the language used; the following example tweets should help
elucidate this point:

\begin{itemize}
  \item \textit{\#canarycall absolutely delighted with the win :) good
  performance, good result}
  \item \textit{\#yellows almost didn't go today glad i did}
  \item \textit{so glad i chose to come today!\#ncfc}
\end{itemize}

It is difficult for a term-based solution to find any common thread here.
Finding the cause of this event would require contextual knowledge of football
matches, team names and commonly employed aliases. The news linkage algorithm
did initially find a news story for the term ``joy'' on this date.
\textit{The British Prime Minister ``let out a little cry of joy'' over David 
Bowie
Scottish independence comments} (Telegraph, Feb 24, 2014). The articles 
returned all concerned this story and were found to be closely related, but 
were dropped from the news linkage because they did not match those returned 
from the other search terms. This
highlights the benefits of searching with multiple terms and ensuring that the 
results are related.

The second example is \texttt{ASL-02-04}. This event was due to
increased levels of air pollution observed in London at the beginning of
April, caused by a Saharan dust cloud.
This event had a \madmax of 12 indicating a significant increase in baseline
activity for the alert period. It was well summarised by all aspects of our
situational awareness algorithm. The top ranked tweets provided by our
summary method (STT) produced tweets more representative of the event than those
from all tweets in the gist. This is demonstrated by the top tweet
selected by both:

\begin{itemize}
  \item STT top tweet: \textit{i can't breathe \#asthma \#smog}
  \item GTT top tweet: \textit{my asthma is literally so bad}
\end{itemize}

Here selecting the top tweets from the filtered event set captures tweets
representative of the event as opposed to the baseline illness activity. The 
news linkage for this example worked well, with all five of the top
selected articles being representative of the event. The top article,
\textit{``Air pollution reaches high levels in parts of England - BBC''}, gives
the cause of the event in the first few lines: \textit{``People with health
problems have been warned to take particular care because of the pollution - a
mix of local emissions and dust from the Sahara.''}

The third case is \texttt{VOL-20-04}. Reading the tweets makes it
clear that this one day event is caused by people feeling sick after
eating too much chocolate on Easter Sunday. In this case the TNT summary and
all tweet ranking return similar tweets as there is little baseline activity and
that baseline activity is not strongly related. The top tweets from both sets
therefore both produce good summaries:

\begin{itemize}
  \item STT top tweet: \textit{seriously i feel sick having all this chocolate}
  \item GTT top tweet: \textit{eaten too much chocolate feel sick}
\end{itemize}

While the top ranked tweets are similar the event tweet filtering does remove
baseline tweets referring to general illness. No good news searches were found
in this case. This event may be
valid in the context of social media but it is not newsworthy. 

The fourth example is \texttt{SAL-11-08} which is the UK Twitter
reaction to the death of Robin Williams. These tweets from the sadness
keyword group exhibit both the highest \madmax (20) and the highest overall
tweet count for any single event (4472). The prominence of celebrity deaths 
within our detected events mirrors earlier findings 
\cite{petrovic2010streaming}. As with all of our high \madmax events the TNT 
tweet ranking and news linkage work well. The top news article returned is an 
article reporting the death of Mr. Williams: \textit{``Robin Williams dies aged 
63 in
suspected suicide''} (Telegraph, August 12, 2014). The top five ranked tweets by
TNT tweet filtering are better than those ranked on all tweets as they remove 
baseline general sadness tweets from the ranking:

\begin{itemize}
  \item STT top tweet: \textit{rip robin williams. sad day}
  \item GTT top tweet: \textit{yep , very sad}
\end{itemize}

\section{Conclusion}
\label{sec:conclusion}

We have presented techniques for event detection and situational awareness based
on Twitter data. We have shown that they are robust and generalisable to
different event classes. New event classes could be added to this system simply
by producing a list of keywords of interest and an optional noise filter.
Our event detection is based on the EARS
bio-surveillance algorithm with a novel filtering mechanism. The maximum Median
Absolute Deviations from the median provides a robust statistic for determining
the strength of relative spikes in count-based time series. As it is based on 
the
median, this measure handles cases where data is non-normal as was the case for
some of our symptom based geo-tagged tweets. The event detection
approach achieved an F1 score of 0.9362 on our event examples.

By filtering to terms that are significantly different ($\alpha < 0.05$) in
frequency from baseline levels we have extracted terms to search news sources
for related articles. Where good news matches are found these revise our event
term list. We have created two novel algorithms that provide additional
situational awareness about an event from these event terms.

Firstly, we rank the filtered set of news articles to produce the top five
representative articles. The news linkage, weighted towards precision, achieved
an F0.5 score of 0.79 on our example set, with no false positives.

Secondly, we produce a top five ranked list of tweets that summarise an event.
These ranked tweets are calculated from the tweet set, filtered by those that
contain the extracted event terms. The top ranked tweets fully matched our
human-coded event summaries in 21 out of 26 cases.

In future work we aim to improve our news linkage algorithm with a final
step checking whether the articles returned are similar to the event tweets, using
cosine similarity or other features such as entities identified in the news
articles. Additional improvements to event detection would lie in improving spam
detection and adding sentiment classification to our emotion example as a
classifier. Collecting data over longer time periods would also allow us to look into
using bio-surveillance algorithms which require seasonal baseline information.

\section{Acknowledgements}

This research was carried out in cooperation with the UK Defence Science and 
Technology Laboratory. It was funded by the U.S. Department of Defense's 
Defense Threat
Reduction Agency (DTRA), through contract HDTRA1-12-D-0003-0010.

\fontsize{9.5pt}{10.5pt} \selectfont
\bibliographystyle{aaai}
\bibliography{ewd}

\end{document}